# Co-propagation of 6 Tb/s (60*100Gb/s) DWDM & QKD channels with ~17 dBm aggregated WDM power over 50 km Standard Single Mode Fiber

P. Gavignet [(1)], F. Mondain [(1)], E. Pincemin [(1)], A. J. Grant [(2)], L. Johnson [(2)], R. I. Woodward [(2)], J. F. Dynes [(2)] and A. J. Shields [(2)]

*(1)   Orange Innovation, 2 Avenue Pierre Marzin, 22300 Lannion, France*
*(2)   Toshiba Europe Ltd, Cambridge, UK*
paulette.gavignet@orange.com

**Abstract:** We report the co-propagation, over 50km of SSMF, of the quantum channel (1310nm) of a QKD system with ~17dBm total power of DWDM data channels (1550nm range). A metric to evaluate Co-propagation Efficiency is proposed.

## 1. Introduction

Several decades have elapsed since Quantum Key Distribution (QKD) [1], was proposed to improve the security of communication networks and to prepare for the future Quantum Internet. Many solutions have been tested in laboratories, some of them in field trials and very few have been implemented in operational networks. However, a large deployment of this new technology by operators still faces some reluctance like the cost and constraints of introduction of QKD systems in the operators' networks. In order to reduce the cost of introduction of QKD in already deployed networks, the co-propagation, on a same fiber, of the quantum channel and existing WDM (Wavelength Division Multiplexing) channels is of utmost importance. But the engineering rules must be carefully chosen so that both types of signals can coexist in the same fiber. This co-propagation can often reduce the Secure Key Rate (SKR) which is particularly sensitive to the length of the transmission link, through an increased Quantum Bit Error Rate (QBER), principally due to Raman scattering of co-propagating light.

In this paper, we report very high SKRs (compared to previous work [2]), with classical & quantum co-existence on several tens of kilometers of Standard Single Mode Fiber (SSMF). Very high aggregated power (~17 dBm) of 100 Gb/s DWDM channels is accepted by the QKD system which makes this configuration fully compatible with currently deployed WDM transmission links. A metric to evaluate the Co-propagation Efficiency (CE) is also proposed to ease the performance comparison of classical & QKD channels co-propagation experiments.

## 2. Description of the QKD system

We use a Multiplexed QKD System, designed and built by Toshiba. This system implements an efficient BB84 QKD protocol with decoy states [3]. Phase-encoded quantum states are generated by a gain-switched laser followed by an asymmetric Mach-Zehnder interferometer in the QKD transmitter, and avalanche photodiodes (APDs) are used in the receiver to measure received states [4]. The total system size is 3 rack-units (3U) per node.

The QKD system is optimized for supporting co-propagation of quantum and classical signals. This is achieved using a quantum channel at 1310 nm, for increased spectral separation from the QKD service channel / multiplexed data channels in the telecom C-band, to minimize the impact of Raman scattering. In addition, the system employs high-extinction spectral filtering and time-domain gating at the QKD receiver to isolate the quantum channel from co-propagating / Raman-scattered light with maximal signal-to-noise ratio. The QKD system also includes automatic self-optimization routines to dynamically adjust various optical parameters for maximal performance on each communication link. Add/drop multiplexing hardware is also included in the QKD Unit to multiplex the quantum channel, QKD service channels and any auxiliary data channels onto a fiber pair communication channel.

## 3. Experimental evaluation of co-propagation

The experimental set-up (Fig. 1) used for the co-propagation evaluation is constituted of a DWDM comb of sixty DP-QPSK (Dual Polarization Quadrature Phase Shift Keying) channels at 100 Gb/s ranged from 1533.6 to 1557 nm (with 50 GHz spacing) and sent to the Aux input (Rx) of the Alice side terminal. This optical comb is based on the interleaving of two amplified DWDM combs (with EDFAs (Erbium Doped Fiber Amplifiers)) of 30 (100 GHz spaced) channels each. Another EDFA is placed after the interleaver leading to up to 18.5 dBm of WDM total input power at the Alice Aux Rx. In that case the WDM total power in the fiber (at the QKD Tx port) is 16.8 dBm due to an insertion loss of 1.7 dB on the path between Aux Rx and QKD Tx (see Fig. 1). An optical attenuator allows to change the WDM aggregated power co-propagating with the quantum channel at 1310 nm. Various spools of SMF-28 single mode fiber have been used (from 20 to 70 km) and a tap coupler (10/90) is inserted after it to perform power or spectra measurements. In order to maximize the launched power for co-propagation, only the Alice to Bob direction is fed with the WDM comb. To further increase the power per channel considered, half of the comb can be removed. This will be indicated when appropriate. The spectra are represented, and we see that in addition to the quantum channel (not on the spectra), the DWDM channels co-propagate with 2 service channels (C59 & C60).

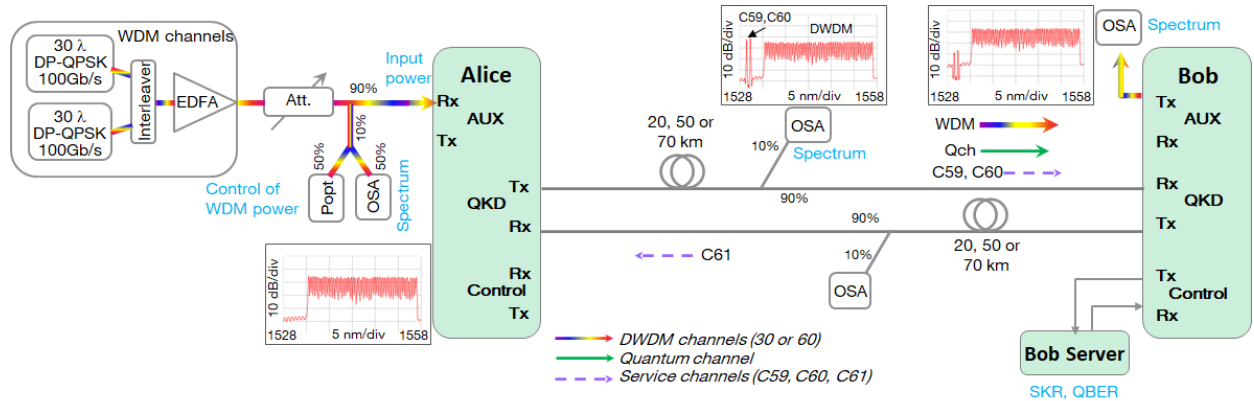

Fig. 1. Experimental set-up for co-propagation evaluation.

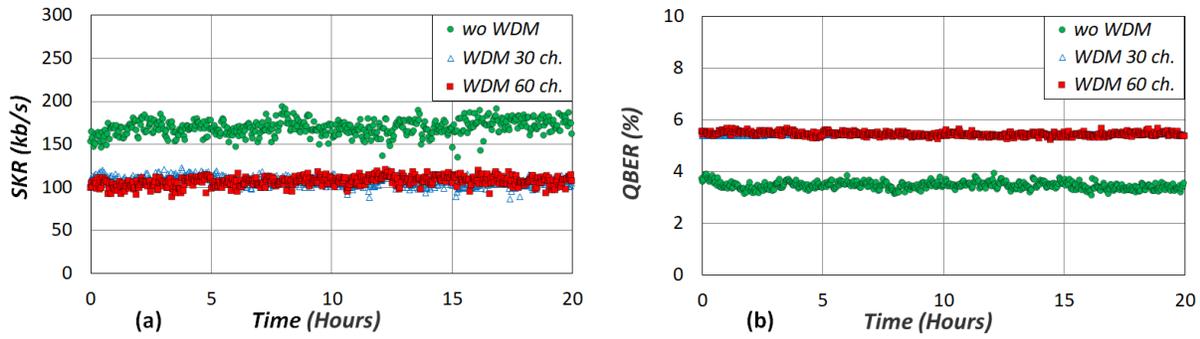

Fig. 2. SKR (a) and QBER (b) with 50 km fiber and 16.8 dBm WDM total power in the fiber.

Fig. 2 shows the SKR (a) and the QBER (b) versus elapsed time for 50 km fiber length for a WDM total power of 16.8 dBm which is the maximum value allowed by the set-up. Green points correspond to the absence of WDM, blue triangles are with 30 channels and red squares are for 60 channels. The mean SKRs with 0, 30 or 60 channels are respectively 169, 107 and 106 kb/s. The corresponding mean QBERs are 3.4, 5.4 and 5.4 %.

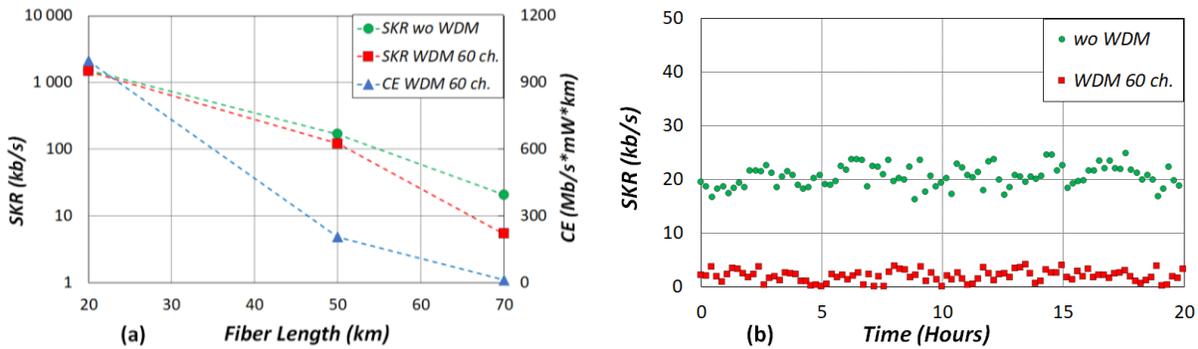

Fig. 3 (a) SKR (left axis) and CE (right axis) versus fiber length for a constant 15.3 dBm WDM total power, (b) SKR with 70 km fiber and 15.8 dBm WDM total power.

We also performed the tests with 20 and 70 km and Fig. 3a shows the evolution, versus the fiber length, of the SKR (left axis) and the Co-propagation Efficiency (CE) (right axis) (that will be introduced in paragraph 4), for a constant WDM total power of 15.3 dBm in the fiber that ensures a fair comparison between the various fiber lengths under study and also allows to protect the Rx from damages at 20 km (where the link losses are very low). For a fiber length of 20 km, the average SKR recorded for 64 hours is equal to 1.47 Mb/s with 60 channels and 15.3 dBm WDM total power. Fig. 3b shows the SKR results in the 70 km configuration with the maximum acceptable aggregated DWDM power coupled in the fiber which is 15.8 dBm. Note that the total link loss with 70 km between Alice and Bob (from QKD Tx to QKD Rx) is 17.5 dB at 1550 nm and 25.7 dB at 1310 nm.

## 4. Analysis and comparison with previous work

The main goal of these tests was to evaluate the possible deployment of QKD system on a WDM link that is already deployed in the field, as well as the number of WDM channels that could be mixed with the QKD signal in the same SMF-28 fiber (to allow DWDM & QKD channels co-existence). If we consider the results of Fig. 2a with 50 km, we see that the SKR is higher than 100 kb/s and the results are the same in the two configurations (30 and 60 channels); this is also the case for the QBER (Fig. 2b). Moreover, we can see that the SKR and QBER values are very stable during time. We have presented results for a period of about 20 hours, but measurements have been performed for several days and confirm the stability over time. The power per channel is ~ +2 dBm in the 30 channels configuration and ~ -1 dBm with 60 channels which is in the range of the power per channel currently used in DWDM 100 Gb/s systems [5]. References [5] and [6] show that a power per channel in the range [0-3]-dBm is optimal for WDM transmission at 100 or 400 Gb/s, making us confident in the co-propagation of several tens of WDM channels with the quantum channel in real field environment. We also performed tests feeding the Bob to Alice direction with the DWDM channels and confirmed there was no modification of the SKR and QBER.

We also notice in Fig. 2 and Fig. 3 that the results in terms of SKR and QBER only depend on the aggregated WDM power. This shows that the number of channels and/or the total bit rate that co-propagate with the quantum channel do not constitute the best parameters to consider, when evaluating the ability of co-existence of classical & QKD channels. Indeed, these experiments confirm that the main parameter to consider is the WDM total power that co-propagates with the quantum channel; the number of channels and the total bit rate are deduced from the tolerated WDM total power.

Thus, we propose a new figure of merit called CE (for Co-propagation Efficiency) to evaluate the performance of classical & QKD signals co-propagation. The proposed CE metric is equal to the following product SKR $*$ $P_{WDM}$ $*$ L (in Mb/s$*$mW$*$km) with respectively SKR, the Secure Key Rate, $P_{WDM}$, the co-propagating classical channels total power, and L, the link length. Indeed, $P_{WDM}$ is the accepted total power and SKR and L refers directly to the ability of the QKD system to deliver a key in a defined configuration. The advantage of this parameter is to be independent from the wavelength of the quantum channel. In our case, we obtain a value of CE of 253.7 Mb/s$*$mW$*$km in the 50 km configuration with 16.8 dBm aggregated DWDM power which is much higher than in reference [2] for which the CE is 9.3 Mb/s$*$mW$*$km. Fig. 3a (right axis) shows the very fast degradation of the CE with fiber length and thus directly reflects the ability of a QKD system to tolerate co-propagation of quantum and classical channels.

## 5. Conclusion-Perspectives

We have reported experimental results of the co-propagation of the quantum channel of a QKD system at 1310 nm with a DWDM comb of 60 channels at 100 Gb/s for a total rate of 6 Tb/s. The emission of a Secure Key is possible with a very high aggregated power: ~17 dBm WDM total power (limited by the set-up) for 50 km SSMF and ~16 dBm for 70 km. These results show the possibility to deploy commercial QKD system on currently existing fully filled WDM links with 100 Gb/s and 400 Gb/s channels in Data Center Interconnection (DCI) applications. We have also proposed a new figure of merit to evaluate the Co-propagation Efficiency and, based on this metric, the results presented here outperforms the previous co-propagating results in close conditions.